\documentclass[aps,prd,twocolumn,showpacs,groupedaddress,floatfix]{revtex4}
\usepackage{graphicx} 
\usepackage{dcolumn}  
\usepackage{bm}       
\usepackage{amssymb}  
\usepackage{multirow}

\begin{document}

\newcommand{\dzero}     {D0}
\newcommand{\ttbar}     {\mbox{$t\bar{t}$}}
\newcommand{\bbbar}     {\mbox{$b\bar{b}$}}
\newcommand{\ccbar}     {\mbox{$c\bar{c}$}}
\newcommand{\herwig}    {\sc{herwig}}
\newcommand{\pythia}    {\sc{pythia}}
\newcommand{\vecbos}    {\sc{vecbos}}
\newcommand{\alpgen}    {\sc{alpgen}}
\newcommand{\singletop}   {\sc{singletop}}
\newcommand{\qq}        {\sc{qq}}
\newcommand{\evtgen}    {\sc{evtgen}}
\newcommand{\tauola}    {\sc{tauola}}
\newcommand{\geant}     {\sc{geant}}
\newcommand{\metcal}    {\mbox{$\not\!\!E_{Tcal}$}}
\newcommand{\met}       {\mbox{$\not\!\!E_T$}}
\newcommand{\pt}        {$p_T$}
\newcommand{\sigmatt}   {\ensuremath{\sigma_{t\bar{t}}}}
\newcommand{\mtop}      {$m_{\rm top}$}
\newcommand{\rsigma}    {\ensuremath{R_{\sigma}}}

\newcommand{\lumi}      {1~$\rm fb^{-1}$}
\newcommand{\result}    {X.X}
\newcommand{\erstat}    {^{+X.X}_{-X.X}}
\newcommand{\ersyspos}  {+X.X}
\newcommand{\ersysneg}  {-X.X}
\newcommand{\ersys}     {^{\ersyspos}_{\ersysneg}}
\newcommand{\erlumi}    {X.X}
\newcommand{\ljets}     {\ensuremath{\ell}+\rm{jets}}
\newcommand{\eplus}     {\ensuremath{e}+jets}
\newcommand{\muplus}    {\ensuremath{\mu}+jets}
\newcommand{\ltau}      {\ensuremath{\tau \ell}}
\newcommand{\etau}      {\ensuremath{\tau e}}
\newcommand{\mutau}     {\ensuremath{\tau \mu}}

\newcommand{\rs}        {XX.X}
\newcommand{\rserr}     {^{+X.X}_{-X.X}}
\newcommand{\rsigmaell}    {\ensuremath{R^{\ell\ell/\ell j}_{\sigma}}}
\newcommand{\rsigmatau}    {\ensuremath{R^{\ltau/\ell \ell\text{-}\ell j }_{\sigma}}}

\newcommand{\mnlo}     {165.5}
\newcommand{\ermnlo}   {^{+6.1}_{-5.9}}
\newcommand{\mtm}      {169.1}
\newcommand{\ermtm}    {^{+5.9}_{-5.2}}
\newcommand{\mtc}      {167.5}
\newcommand{\ermtc}    {^{+5.8}_{-5.6}}
\newcommand{\mtk}      {168.2}
\newcommand{\ermtk}    {^{+5.9}_{-5.4}}

\hspace{5.2in} \mbox{FERMILAB-PUB-09-092-E}

\title{Combination of $t\bar{t}$ cross section measurements and constraints on the mass of the top quark and its decays into charged Higgs bosons}

%
\author{V.M.~Abazov$^{37}$}
\author{B.~Abbott$^{75}$}
\author{M.~Abolins$^{65}$}
\author{B.S.~Acharya$^{30}$}
\author{M.~Adams$^{51}$}
\author{T.~Adams$^{49}$}
\author{E.~Aguilo$^{6}$}
\author{M.~Ahsan$^{59}$}
\author{G.D.~Alexeev$^{37}$}
\author{G.~Alkhazov$^{41}$}
% ...

\affiliation{\vspace{0.1 in}(The D\O\ Collaboration)\vspace{0.1 in}}
% ...
\affiliation{$^{6}$University of Alberta, Edmonton, Alberta, Canada;
                Simon Fraser University, Burnaby, British Columbia, Canada;
                York University, Toronto, Ontario, Canada and
                McGill University, Montreal, Quebec, Canada}
% ...
\affiliation{$^{30}$Tata Institute of Fundamental Research, Mumbai, India}
% ...
\affiliation{$^{37}$Joint Institute for Nuclear Research, Dubna, Russia}
% ...
\affiliation{$^{41}$Petersburg Nuclear Physics Institute,
                St. Petersburg, Russia}
% ...
\affiliation{$^{49}$Florida State University, Tallahassee, Florida 32306, USA}
% ...
\affiliation{$^{51}$University of Illinois at Chicago,
                Chicago, Illinois 60607, USA}
% ...
\affiliation{$^{59}$Kansas State University, Manhattan, Kansas 66506, USA}
% ...
\affiliation{$^{65}$Michigan State University,
                East Lansing, Michigan 48824, USA}
% ...
\affiliation{$^{75}$University of Oklahoma, Norman, Oklahoma 73019, USA}
% ...

\date{June 3rd 2009}

\begin{abstract}
We  combine measurements of the  top quark pair production cross section 
 in $p\bar{p}$ collisions in the $\ell$+jets, $\ell\ell$ 
and \ltau\ final states  (where $\ell$ is an electron or muon) at a center of mass energy of \( \sqrt{s}=1.96
 \)~TeV in $1$~fb$^{-1}$ of data collected with the D0 detector.
For a top quark mass of 170~GeV/$c^2$, we  obtain $\sigma_{t\bar{t}}=8.18^{+0.98}_{-0.87}$~pb 
in agreement with the theoretical prediction. 
Based on
 predictions from higher order quantum chromodynamics, we extract 
a mass for the top quark from the combined \ttbar\ cross section, consistent 
with the world average of the top quark mass.
In addition, the ratios of \ttbar\ cross sections in different final states are used to set upper
limits on the branching fractions $B(t\rightarrow H^{+}b \rightarrow
 \tau^{+}\nu b)$ and $B(t\rightarrow H^{+}b \rightarrow c\bar{s} b)$ as a
 function of charged Higgs boson mass. 
\end{abstract}

\pacs{12.15.Ff, 13.85.Lg, 13.85.Qk, 13.85.Rm, 14.65.Ha, 14.80.Cp}

\maketitle

Precise measurements of the production and decay 
properties of the heaviest known fermion, the top quark,   
provide important tests of the standard model (SM) and offer a window for
searches for new physics. In this paper we
measure the top-antitop quark pair (\ttbar) production cross
section and compare it with the SM prediction, extract the top quark
pole mass from this measurement and search for new physics in top
quark decays analyzing ratios of the \ttbar\ cross sections measured in
different decay channels.

The inclusive \ttbar\ production cross section
(\sigmatt) is measured in different \ttbar\ decay
channels assuming SM branching fractions. The comparison of the results to 
predictions in next-to-leading order perturbative quantum chromodynamics (QCD), including higher order soft gluon resummations  
~\cite{nadolsky,cacciari,moch,kidonakis}, yields a direct test of the SM. Ratios of \sigmatt\ measured in
different final states are particularly sensitive to non-SM particles that may appear in top quark decays, especially if the boson in the decay is not a SM $W$ boson.  An example is the decay into  a charged
Higgs boson ($t\rightarrow H^+ b$), which, as predicted in some models~\cite{guasch}, 
can compete with the SM decay $t\rightarrow W^+ b$. Additionally, many experimental uncertainties cancel in the ratios. 
Furthermore, since \sigmatt\ depends on the mass of the top quark ($m_{t}$),
it can be used to extract $m_{t}$. Such measurement is less accurate than
direct mass measurements, but provides complementary information with different experimental and theoretical uncertainties. 
 
Within the SM,  each quark of the \ttbar\ pair
is expected to decay nearly 100\% of the times into a $W$~boson
and a $b$~quark~\cite{Amsler:2008zzb}.
$W$ bosons can decay hadronically into
$q\bar{q}^\prime$ pairs or 
leptonically into $e\nu_{e}$, $\mu\nu_{\mu}$ and $\tau\nu_{\tau}$   with
              the $\tau$ in turn decaying onto an electron, a muon, or  hadrons,
              and associated neutrinos.   
If one of the $W$ bosons decays hadronically while the other one produces  
a direct electron or muon  or a secondary electron or muon from  $\tau$ decay, the final state is 
referred to as the $\ell$+jets (or $\ell$j) channel.
If both $W$ bosons  decay leptonically, this leads to a dilepton final state containing a pair of electrons, a pair of muons, or 
an electron and a muon (the $\ell \ell$ channel), or a hadronically decaying
tau accompanied either by an electron or a muon (the \ltau\ channel). 

Measurements of the individual \ttbar\ cross sections in $\ell \ell$ and   
\ltau\ channels using about $1$~fb$^{-1}$ of $p\bar{p}$ data from the D0 detector at the Fermilab Tevatron collider at $\sqrt{s}=1.96$~TeV 
are available in Ref.~\cite{dilepton}.
In the \ljets~channel, we use the same  selection and background estimation as in Ref.~\cite{ljets}, but a slightly larger dataset and a unified treatment of systematic uncertainties with the $\ell \ell$ and \ltau\ channels. 
We provide a brief summary of the event selection and analysis procedures below.  

In each final state we select data samples enriched in $t\bar{t}$ events by
requiring one  or two  isolated high transverse momentum ($p_T$) leptons for the
\ljets\ or $\ell \ell$  channel respectively.  At least two  high $p_T$ jets
are required for $\ell \ell$ and \ltau\ events, and at least three for 
\ljets\ events.  Further, in all but
the $e\mu$ channel, large transverse missing energy (\met)  is required to
account for the large transverse momenta of neutrinos from $W$ boson or $\tau$
lepton decays. In the $e\mu$  final state, a requirement on the sum of the $p_T$
of the highest $p_T$ (leading) lepton and the two leading  jets is imposed instead.   
In the $\mu\mu$ channel, the \met\ requirement is supplemented with  a requirement on the significance of the \met\
measurement, estimated from the $p_T$ of muons and jets, and their expected
resolutions. 
Additional  criteria are applied on the invariant mass of the two opposite 
charge leptons of the same flavor in the $ee$ and $\mu\mu$ channels to reduce the dominant  
background from $Z/\gamma^{\ast} \rightarrow \ell^+ \ell^-$ events.  
In the \ljets\ and \ltau\ channels we require a minimum azimuthal angle separation 
between the \met\ vector and the
 lepton $p_T$, $\Delta\phi(\ell,\met)$, to reduce background from multijet events, where jets are misidentified as electron, muon or $\tau$. Details of lepton, jet and \met\ identification are provided in Refs.~\cite{p14topo,p14dilep}. 
 The final 
 selection in these channels demands at least one identified $b$~jet via a
 neural-network based algorithm~\cite{btagging}.  In the \ljets~channel we separate events with one or $\ge2$ b-tagged jets due to their different signal over background ratio and systematic uncertainties. 

\begin{table*}[ht]
\begin{center}
\caption{Expected numbers of background and signal events for $\sigma_{t\bar{t}}=8.18$~pb, observed numbers of data events and measured $\sigma_{t\bar{t}}$ at top mass of $170$~GeV/$c^2$. Quoted
  uncertainties include both statistical and systematic uncertainties, added
  in quadrature. }
\begin{tabular}{l|c|ccccc|c|c|c}\hline \hline
Channel & Luminosity(pb$^{-1}$)&$W$+jets & $Z$+jets & Multijet   & Other bkg& $t\bar{t}$ & Total  & Observed & $\sigma_{t\bar{t}}$~(pb)   \\[2pt] \hline
$e$+jets (3 jets, 1 $b$~tag) &  1038 & 53.4$^{+6.0}_{-6.0}$ & $\phantom{0}$6.0$^{+1.2}_{-1.2}$  & 31.5$^{+3.5}_{-3.5}$    & 11.4$^{+1.5}_{-1.4}$    & 81.7$^{+6.4}_{-6.7}$ & 184.0$^{+9.0}_{-9.2}$& 183 & $\phantom{0}8.06^{+1.89}_{-1.71}$ \\[2pt] 
$\mu$+jets (3 jets, 1 $b$~tag) &  996 & 59.2$^{+5.5}_{-5.6}$ & $\phantom{0}$6.5$^{+1.3}_{-1.3}$  & $\phantom{0}$9.7$^{+2.8}_{-2.8}$    & $\phantom{0}$9.5$^{+1.2}_{-1.2}$    & 59.0$^{+5.7}_{-5.6}$ & 143.9$^{+8.1}_{-8.1}$& 133 & $\phantom{0}6.43^{+2.22}_{-2.01}$ \\[2pt] \hline 
$e$+jets (3 jets, $\ge2$~$b$~tags) & 1038 & $\phantom{0}$5.0$^{+0.8}_{-0.8}$  & $\phantom{0}$0.6$^{+0.2}_{-0.2}$      & $\phantom{0}$2.7$^{+0.3}_{-0.3}$
& $\phantom{0}$2.4$^{+0.4}_{-0.4}$     & 30.7$^{+3.9}_{-3.9}$& $\phantom{0}$41.5$^{+4.7}_{-4.6}$ &
40 & $\phantom{0}7.78^{+2.41}_{-2.01}$\\[2pt] 
$\mu$+jets (3 jets, $\ge2$~$b$~tags) & 996 &$\phantom{0}$5.8$^{+0.9}_{-0.9}$  & $\phantom{0}$0.7$^{+0.2}_{-0.2}$      & $\phantom{0}$1.0$^{+0.3}_{-0.3}$
& $\phantom{0}$2.1$^{+0.3}_{-0.3}$     & 23.8$^{+3.4}_{-3.2}$& $\phantom{0}$33.5$^{+4.1}_{-3.9}$ &
31 & $\phantom{0}7.29^{+2.73}_{-2.25}$\\[2pt] \hline 
$e$+jets ($\ge 4$ jets, 1 $b$~tag)  & 1038  & $\phantom{0}$8.5$^{+2.7}_{-2.7}$   & $\phantom{0}$2.2$^{+0.5}_{-0.5}$      & $\phantom{0}$7.9$^{+1.0}_{-1.0}$     & $\phantom{0}$3.0$^{+0.5}_{-0.5}$    & 81.6$^{+8.7}_{-9.1}$& 103.3$^{+7.3}_{-7.6}$ & 113 & $\phantom{0}9.38^{+1.82}_{-1.52}$ \\[2pt] 
$\mu$+jets ($\ge 4$ jets, 1 $b$~tag)  &  996 & 13.6$^{+2.6}_{-2.7}$   & $\phantom{0}$2.5$^{+0.7}_{-0.6}$      & $\phantom{0}$0.0$^{+0.0}_{-0.0}$     & $\phantom{0}$2.4$^{+0.4}_{-0.4}$    & 65.9$^{+6.9}_{-7.2}$& $\phantom{0}$84.3$^{+5.9}_{-6.3}$ & 99 & $10.44^{+2.11}_{-1.76}$\\[2pt] \hline 
$e$+jets ($\ge 4$ jets, $\ge2$~$b$~tags) &  1038 & $\phantom{0}$1.0$^{+0.3}_{-0.3}$   & $\phantom{0}$0.2$^{+0.1}_{-0.1}$       & $\phantom{0}$1.1$^{+0.1}_{-0.1}$
& $\phantom{0}$0.9$^{+0.2}_{-0.2}$     & 41.7$^{+6.0}_{-6.0}$ & $\phantom{0}$44.9$^{+6.0}_{-6.0}$ &
30 & $\phantom{0}5.12^{+1.59}_{-1.28}$ \\[2pt] 
$\mu$+jets ($\ge 4$ jets, $\ge2$~$b$~tags) & 996 & $\phantom{0}$1.5$^{+0.4}_{-0.4}$   & $\phantom{0}$0.3$^{+0.1}_{-0.1}$       & $\phantom{0}$0.0$^{+0.0}_{-0.0}$
& $\phantom{0}$0.7$^{+0.1}_{-0.1}$     & 35.6$^{+5.0}_{-5.1}$ & $\phantom{0}$38.2$^{+5.1}_{-5.2}$ &
34 & $\phantom{0}7.60^{+2.11}_{-1.70}$\\[2pt] \hline 
$ee$                & 1074 &                  & $\phantom{0}$2.3$^{+0.5}_{-0.5}$        & $\phantom{0}$0.6$^{+0.4}_{-0.4}$ 
& $\phantom{0}$0.5$^{+0.1}_{-0.1}$    & 11.6$^{+1.2}_{-1.2}$ & $\phantom{0}$15.0$^{+1.5}_{-1.5}$ &
17 & $\phantom{0}9.61^{+3.47}_{-2.84}$ \\[2pt] 
$e\mu$ (1 jet)                &  1070 &                    & $\phantom{0}$5.5$^{+0.7}_{-0.8}$
& $\phantom{0}$0.9$^{+0.3}_{-0.2}$  & $\phantom{0}$3.1$^{+0.7}_{-0.7}$    & $\phantom{0}$8.9$^{+1.4}_{-1.4}$ & $\phantom{0}$18.4$^{+1.9}_{-1.9}$ &
21 & $10.61^{+5.33}_{-4.23}$ \\[2pt] 
$e\mu$ ($\ge2$ jets)                & 1070 &                   & $\phantom{0}$5.4$^{+0.9}_{-1.0}$       & $\phantom{0}$2.6$^{+0.6}_{-0.5}$ 
& $\phantom{0}$1.4$^{+0.4}_{-0.4}$    & 36.4$^{+3.6}_{-3.6}$ & $\phantom{0}$45.8$^{+4.5}_{-4.5}$ &
39 & $\phantom{0}6.66^{+1.81}_{-1.52}$ \\[2pt]  
$\mu\mu$                &  1009 &                    & $\phantom{0}$5.6$^{+1.1}_{-1.2}$        & $\phantom{0}$0.2$^{+0.2}_{-0.2}$ 
& $\phantom{0}$0.6$^{+0.1}_{-0.1}$    & $\phantom{0}$9.1$^{+1.0}_{-1.0}$ & $\phantom{0}$15.4$^{+1.8}_{-1.9}$ &
12 & $\phantom{0}5.08^{+3.82}_{-3.06}$ \\[2pt] \hline 
$\tau e$  ($\ge 1$~$b$~tag)         & 1038 & $\phantom{0}$0.6$^{+0.0}_{-0.1}$   & $\phantom{0}$0.6$^{+0.1}_{-0.1}$       & $\phantom{0}$3.0$^{+1.7}_{-1.7}$
& $\phantom{0}$0.2$^{+0.1}_{-0.1}$     & 10.7$^{+1.3}_{-1.3}$ & $\phantom{0}$15.0$^{+2.2}_{-2.2}$&  16 & $\phantom{0}8.94^{+4.03}_{-3.32}$\\[2pt]
$\tau\mu$   ($\ge 1$~$b$~tag)         & 996 & $\phantom{0}$0.8$^{+0.1}_{-0.2}$   & $\phantom{0}$1.2$^{+0.3}_{-0.3}$       & $\phantom{0}$8.0$^{+2.8}_{-2.8}$
& $\phantom{0}$0.2$^{+0.0}_{-0.0}$     & 12.6$^{+1.4}_{-1.4}$ & $\phantom{0}$22.7$^{+3.2}_{-3.2}$&  20 & $\phantom{0}6.40^{+3.88}_{-3.43}$ \\[2pt]
\hline \hline
\end{tabular}
\label{tab:yields}
\end{center}
\end{table*}

To simplify the combination and extraction of cross section ratios,
all channels are constructed to be mutually exclusive.  
In particular, events with two identified leptons are excluded from 
the \ljets\ selection, and 
all \ltau\ candidates are removed from the rest of the channels.

The compositions of the samples in the \ljets, $\ell \ell$ and \ltau\ channels 
are shown in Table~\ref{tab:yields}. $W$+jets production dominates
the background for the  \ljets\ events, while multijet production is the most
important background 
in the \ltau\ channel. Background in the $\ell \ell$ channels comes mainly from 
$Z$+jets production. In the $\ell \ell$ channel, contributions from $W$+jets 
production are part of the multijet background. The smaller contribution from 
diboson production is included in 
the category labeled ``other background''. This category also includes the 
contribution from  single top quark production in the  \ljets\ and \ltau\ channels.
The signal, $W$+jets and $Z$+jets backgrounds are simulated using
{\alpgen}~\cite{alpgen} for the matrix element
calculation and {\pythia}~\cite{pythia}
for parton showering and hadronization. Diboson and single top 
backgrounds are simulated with the
{\pythia} and {\singletop}~\cite{single_top}
generators, respectively. We estimate the multijet background  
from the control data samples.
The difference in the ratio
of \ttbar\ and $W$+jets events in the \eplus\  
and \muplus\ final states is the result of the
larger efficiency and misidentified lepton rate
in the \eplus\ channel compensating for the
lower lepton acceptance ($|\eta|<1.1$) compared to the \muplus\ channel ($|\eta|<2.0$). 
In addition, the wider rapidity distribution of the $W$+jets events 
compared to \ttbar\ events increases the $W$+jets background
contribution in the \muplus\ channel.

To calculate the combined cross section, we define a joint likelihood function
as the product  of
Poisson probabilities for the 14 disjoint subsamples, as listed in Table~\ref{tab:yields}.
Fourteen  additional Poisson terms
constrain the multijet background in the \ljets\ and \ltau\ channels.
In particular, for the \etau\ and \mutau\ channels, the multijet 
background is determined by counting  events with an electron or muon and
associated $\tau$ of the same electric charge, introducing a corresponding
Poisson  term 
per channel. In the \ljets\ channel, we estimate the multijet background 
separately for each of the eight subchannels by using  corresponding 
control data samples~\cite{p14btag}. Four additional terms arise from 
applying  this same method in evaluating the multijet background before 
$b$~tagging.     

Each systematic uncertainty is included in the likelihood function 
through one free ``nuisance'' parameter~\cite{p14btag}. Each of these parameters is represented  
by a Gaussian probability density function with zero mean 
and a standard deviation of one;  
all are allowed to float in the maximization of the  likelihood
function, thereby  changing the central value of the measured \sigmatt. 
Correlations are taken into account by
using the same nuisance parameter for a common source  of 
systematic uncertainty in different channels scaled by the corresponding   
 standard deviation  each individual channel.  
Thus, the likelihood function to be maximized is represented by the product
\begin{equation} 
{\cal L}\! =\! \prod_{i=1}^{14}\! {\cal P}(n_{i}, m_{i}) 
\!\times \!\prod_{j=1}^{14}\!  {\cal P}(n_{j}, m_{j}) 
\!\times\! \prod_{k=1}^{K}\! {\rm SD}_{ik} \times\! {\cal G}(\nu_k;0,1) \,,
\label{eq:mlikeli}
\end{equation}
where ${\cal P}(n,m)$ is the Poisson
probability to observe $n$ events given the expectation of 
$m$ events. The predicted number of events in each channel
is the sum of the predicted background and  expected \ttbar\ events, 
which depends on \sigmatt. 
In the product, $i$ runs over the subsamples and $j$ runs over the multijet 
background subsamples.   
The Gaussian distributions ${\rm SD}_{ik} \!\times\! {\cal G}(\nu_k;0,1)$ describe 
the systematic uncertainties, 
$K$ is the total number of independent sources of systematic uncertainty, 
$\nu_k$ are the individual nuisance parameters, and ${\rm SD}_{ik}$ is one standard 
deviation for the source of uncertainty $k$ in subsample $i$. 

Systematic uncertainties on the measured \sigmatt\ are 
evaluated from sources that include 
electron and muon identification; $\tau$ and jet identification and 
energy calibration; $b$-jet identification; modeling of triggers, signal and
background; and integrated luminosity. 
All these uncertainties are treated as 
fully correlated among channels and between signal and background. 
Systematic uncertainties arising from limited statistics 
of data or Monte Carlo samples used in estimating signal or backgrounds 
are considered to be uncorrelated. A detailed discussion  on systematic
uncertainties can be found in Refs.~\cite{dilepton, ljets}.
Table~\ref{tab:systematics} shows a 
breakdown of uncertainties on the combined cross section. We evaluate the effect from each source 
by setting all uncertainties to zero except the one in question  
and redoing the likelihood maximization with respect to only the corresponding nuisance parameter. Since the method allows each 
uncertainty to change the central value, the total uncertainty on 
\sigmatt\ differs slightly from the quadratic 
sum of the statistical and individual systematic uncertainties. The
total systematic 
uncertainty on \sigmatt\ exceeds the 
statistical contribution. The luminosity uncertainty of 6.1\% which enters into the estimation of 
the majority of the backgrounds and the luminosity measurement of the selected samples  
is the dominant source of systematic uncertainty. 

\begin{figure*}[t]
\begin{center}
\setlength{\unitlength}{1.0cm}
\begin{picture}(18.0,4.0)  
\put(0.0,0.2){\includegraphics[height=4.3cm]{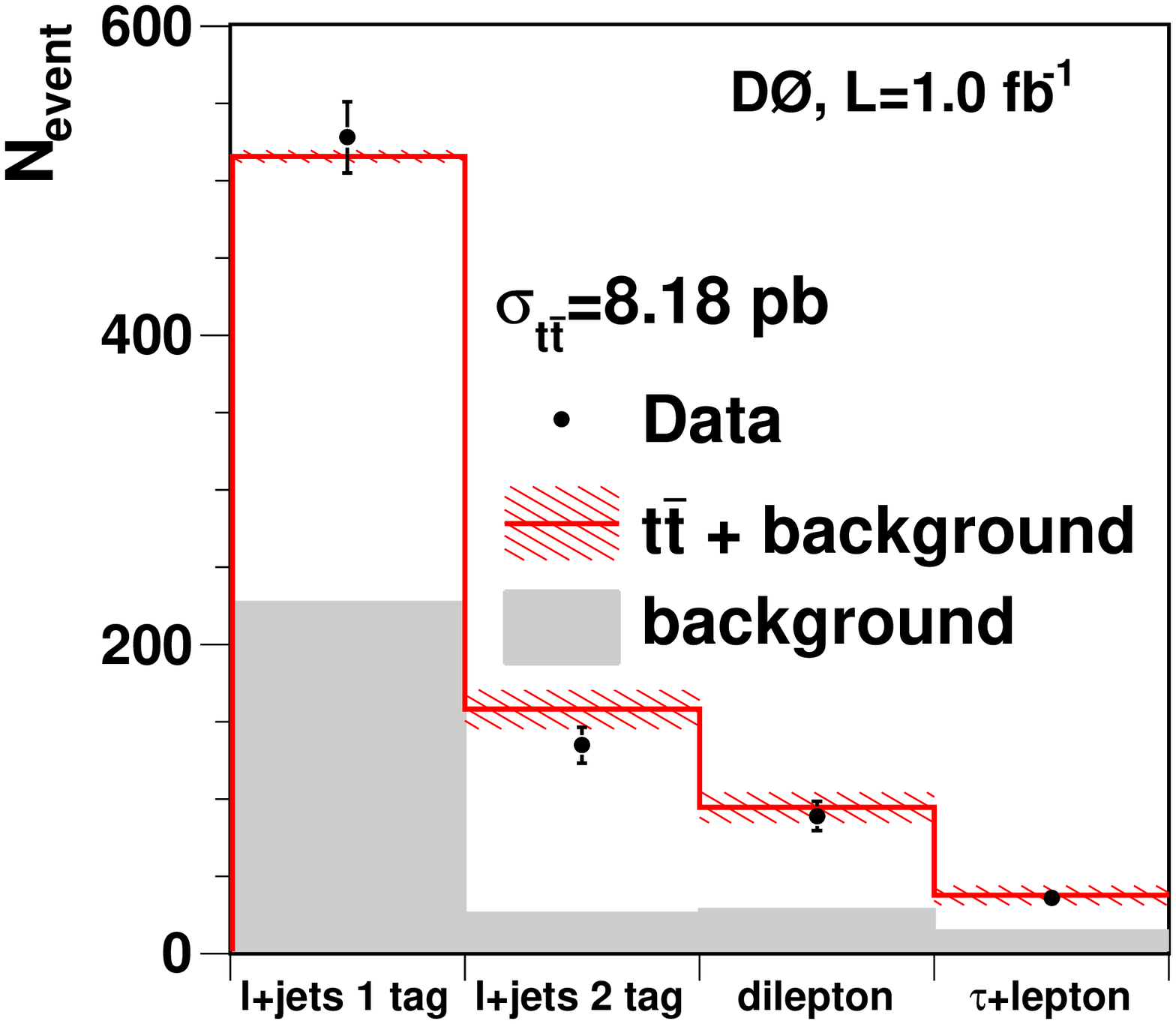} }
\put(4.8,0.2){\includegraphics[height=4.3cm]{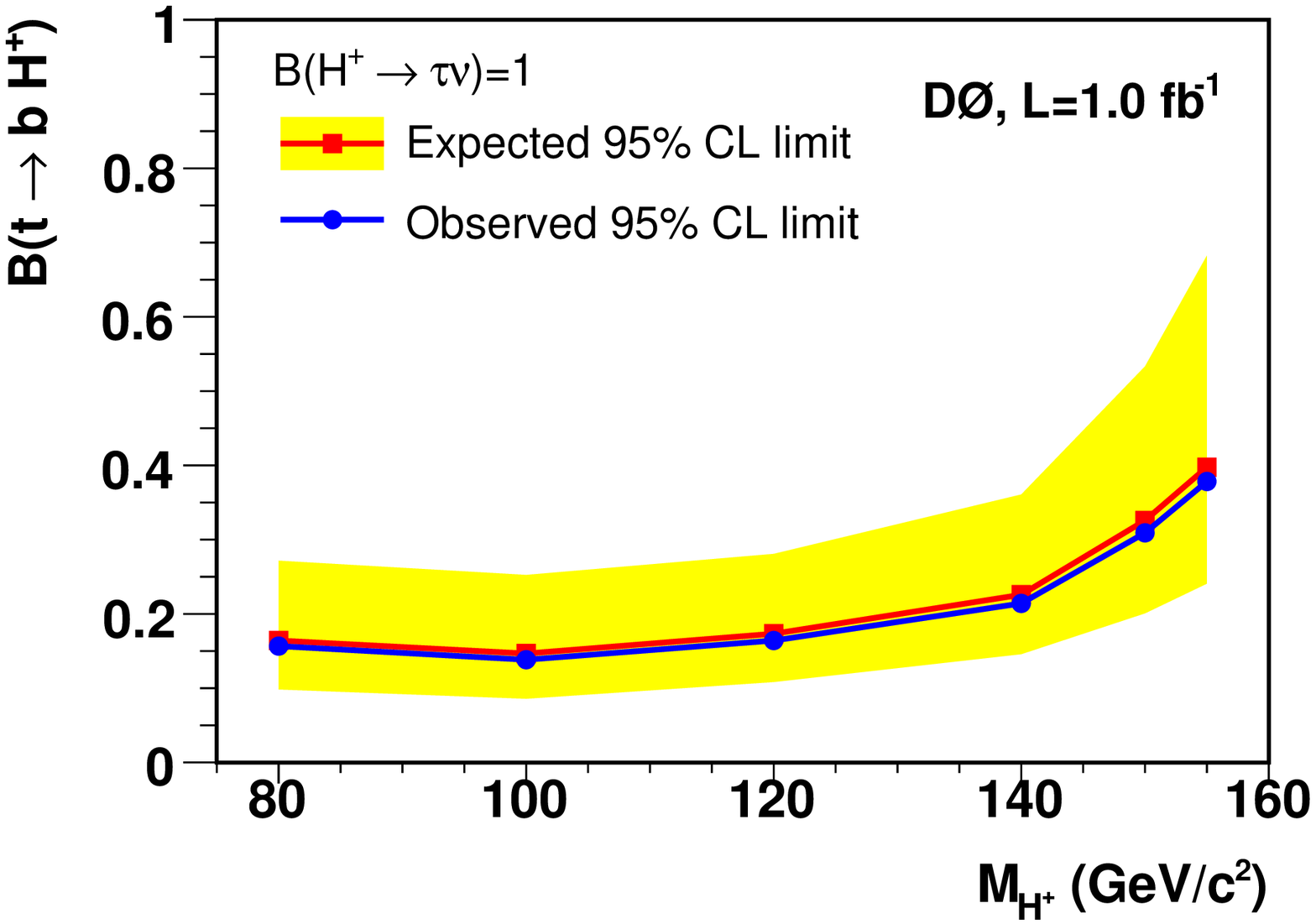} }
\put(11.4,0.2){\includegraphics[height=4.3cm]{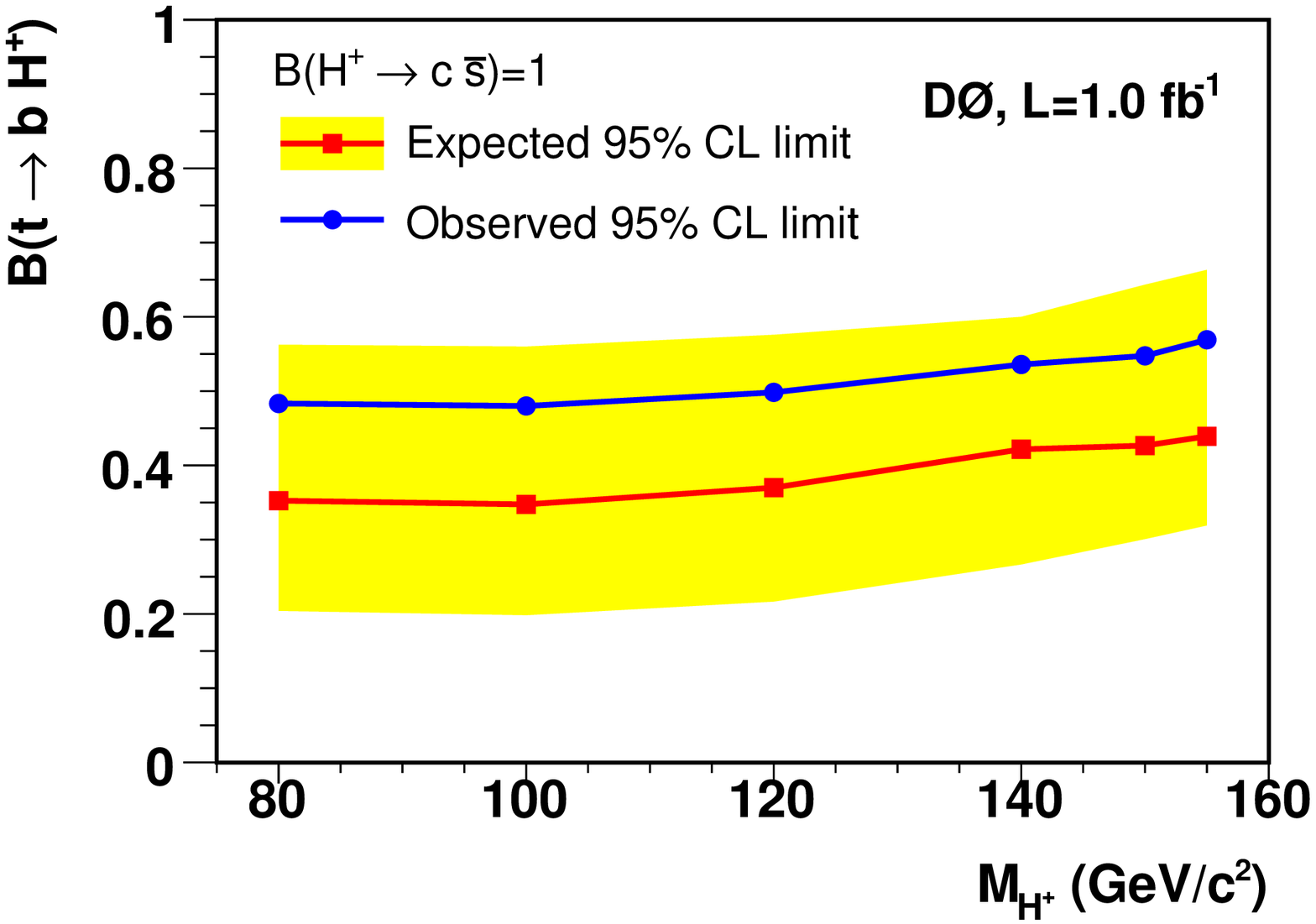} }
\put(3.8,1.34){(a)}
\put(10.3,1.34){(b)}
\put(16.9,1.34){(c)}
\end{picture}
\vspace{-0.8cm}
\caption{(a)~Expected and observed numbers of events versus channel, used in
  measuring the combined \sigmatt. The dashed band around the prediction
  indicates the total uncertainty.  Upper limits on
  $B(t\rightarrow H^{+}b)$ for (b)~tauonic and (c)~leptophobic $H^{+}$
  decays. The yellow band shows the $\pm1$~standard deviation band around the expected limit. }
\label{fig:limits}
\end{center}
\end{figure*}

Table~\ref{tab:result} summarizes the individual \sigmatt\ measurements for
the individual channels, as well as some of their combinations.  
Within uncertainties, all measurements are consistent with each other. 
The combined cross section for  \ljets, $\ell \ell$ and $\ltau$
final states  for a top quark mass of $170$~GeV/$c^2$ is evaluated to be 
\begin{equation}
\sigma_{t\bar{t}}=8.18^{+0.98}_{-0.87}\text{~pb} \,,
\end{equation}
in agreement with theoretical predictions~\cite{nadolsky,cacciari,moch,kidonakis}. The uncertainty is comparable to the one on the cross section combination from different methods in the \ljets\ channel performed by \dzero~\cite{ljets}.
The observed number of events in the different  channels is compared to the sum 
of the background and combined \ttbar ~signal 
in Fig.~\ref{fig:limits}(a).

\begin{table}[t]
\begin{center}
\caption{Summary of uncertainties on  the combined \sigmatt.} 
\vspace{0.2cm}
\begin{tabular}{ccc} 
\hline \hline
 Source  &       \multicolumn{2}{c}{$\Delta\sigmatt$~(pb)  } \\ \hline
                                       Statistical &  +0.47 &   $-0.46$ \\
\hline
                                  Lepton identification & +0.15 &   $-0.14$  \\
                                     Tau identification & +0.02 &   $-0.02$  \\
                                     Jet identification & +0.11 &   $-0.11$  \\
                                        Jet energy scale & +0.19 &   $-0.16$  \\
                                        Tau energy scale &  +0.02 &   $-0.02$  \\
                                                Trigger modeling &  +0.11 &   $-0.07$  \\
                                   $b$ jet identification &  +0.34 &   $-0.32$  \\
                                        Signal modeling & +0.17 &   $-0.15$  \\
                                         Background estimation &  +0.14 &   $-0.14$  \\
                                        Multijet background & +0.12 &   $-0.12$  \\
                                              Luminosity & +0.56 &   $-0.48$  \\
                                                 Other & +0.15 &   $-0.14$  \\
\hline
              Total systematic uncertainty & +0.78 &   $-0.69$   \\
\hline \hline
 \end{tabular}
\label{tab:systematics}
 \end{center}
\end{table}

\begin{table}[t]
\begin{center}
\caption{Summary of  measured \sigmatt\ in different channels for $m_t=170$~GeV/$c^2$.} 
\vspace{0.2cm}
\begin{tabular}{c|c} 
\hline \hline
Channel & \sigmatt~(pb) \\[2pt] \hline \\[-8pt]
\ljets & $8.46^{+1.09}_{-0.97}$ \\[2pt]
$\ell \ell$~\cite{dilepton} & $7.46^{+1.60}_{-1.37}$ \\[2pt]
\ljets\ and $\ell \ell$ & $8.18^{+0.99}_{-0.87}$  \\ [2pt]
\ltau~\cite{dilepton} & $7.77^{+2.90}_{-2.47}$\\ [2pt]
\ljets, $\ell \ell$ and \ltau & $8.18^{+0.98}_{-0.87}$ \\[1pt] 
\hline \hline
\end{tabular}
\label{tab:result}
\end{center}
\end{table}

We compute ratios \rsigma\ of measured cross sections, 
$\rsigmaell= \sigma_{t\bar{t}}^{ \ell \ell}/\sigma_{t\bar{t}}^{\ell\rm{j}}$ and
$\rsigmatau= \sigma_{t\bar{t}}^{\ltau}/\sigma_{t\bar{t}}^{\ell\rm{j} \& \ell \ell}$, by generating pseudo-datasets
 in the numerator and denominator in order to take into account the
 correlation between systematic uncertaintes. 
 $\sigma_{t\bar{t}}^{\rm channel}$ represent the
measured cross sections in the corresponding channel.
The pseudo-datasets are created by varying the number of signal and background
events around the expected number according to Poisson probabilities. All independent sources of systematic uncertainties are varied within a Gaussian distribution. 
Although the  individual channels considered are exclusive,
each channel can receive signal contributions from different
$t\bar{t}$ decay modes.
We calculate the contribution from dilepton events to the \ljets~final
state as well as the contribution from dilepton and \ljets~events to
the \ltau\ final states using the
corresponding observed cross sections in the individual channels when
generating pseudo-datasets.
For each pseudo-dataset, we perform the maximization of
Eq.~\ref{eq:mlikeli} separately in the numerator and denominator, and
divide the results. The central value is obtained from the mode of the
distribution of \rsigma, and the uncertainties are derived from the
interval containing 68\% of the pseudo-experiments.
From these pseudo-experiments we obtain
$\rsigmaell=0.86^{+0.19}_{-0.17}$ and
$\rsigmatau=0.97^{+0.32}_{-0.29}$\,, which is
consistent with the SM expectation of $\rsigma=1$.
  
Extensions of the SM, based on supersymmetry or grand unification~\cite{guasch}, 
require the existence of additional Higgs multiplets beyond the Higgs doublet of the SM. 
Some of these models, such as the Two Higgs-Doublet Model or  
the Minimal Supersymmetric Standard Model, foresee the existence
of physical degrees of freedom which can be associated with a 
charged scalar particle, the charged Higgs boson. 
If this charged Higgs boson is lighter than the top quark, it will 
appear in the top quark decays.
We use the  ratios to extract upper limits on the branching ratio 
$B\equiv B(t\rightarrow H^{+}b)$. 
In particular, a charged Higgs boson decaying
into a tau and a neutrino ($B(H^{+}\rightarrow\tau\nu)=1$) results in more events in the \ltau\ channel, while
fewer events appear in the $\ell\ell$ and \ljets\ final states compared to the
SM prediction. In case of a leptophobic ($B(H^{+}\rightarrow c\bar{s})=1$)
model, the number of dilepton events decreases faster than the number of
\ljets\ events for increasing $B(t\rightarrow H^{+}b)$. We therefore use \rsigmaell\ to set limits 
on the leptophobic model,  
while \rsigmatau\ is used to search for decays in which  the charged Higgs bosons are assumed to decay exclusively to taus. 

To extract the limits, we generate 
pseudo-datasets assuming different branching fractions $B(t\rightarrow
H^{+}b)$. The signal for a charged Higgs boson is
simulated using the 
{\pythia} Monte Carlo event generator~\cite{pythia}, and includes
decays of 
$t\bar{t}\rightarrow W^{+}bH^{-}\bar{b}$ and its charge conjugate ($WH$)
and $t\bar{t}\rightarrow H^{+}bH^{-}\bar{b}$ ($HH$). 
For a given branching fraction $B$, we calculate the expected number of $t\bar{t}$ events per 
final state, 
\begin{equation}
N_{t\bar{t}} \! =\! [ (1\!-\!B\!)^2\! \cdot\! \epsilon_{WW}\! +\! 2B\,(1 \!-\! B\!)
\! \cdot\! \epsilon_{WH} \! + \! B^2 \! \cdot\!  \epsilon_{HH} \!] \sigmatt L \; ,
\end{equation}
where $\epsilon$ are the selection efficiencies for the
 different decays ($WW$ refers to $t\bar{t}\rightarrow W^{+}bW^{-}\bar{b}$) and $L$ is the integrated luminosity. We add $N_{t\bar{t}}$ to the expected background and treat the sum as a new number of expected
 events in each channel.
We then perform the likelihood maximization to extract 
\sigmatt\ from these pseudo-data as if they contained only SM \ttbar\ 
production. This provides distributions for the ratios of cross
 sections  for each generated $B$, which are compared to the observed ratio. We set  limits on $B$ by using the frequentist approach of Feldman and
Cousins~\cite{feldmancousins}.

The observed and expected (i.e., for $\rsigma=1$) limits for the 
tauonic  and the leptophobic charged Higgs boson models 
are shown in Figs.~\ref{fig:limits}(b) and~\ref{fig:limits}(c), respectively. 
In the tauonic model the upper $95\%$ CL limits on $B$ range from 15\%
to 40\% for $80~\text{GeV}/c^2\le M_{H^{+}}\le 155~\text{GeV}/c^2$,
improving the limits given in~\cite{CDF_hp}. For the leptophobic
charged Higgs boson model, which is investigated here for the first
time, the upper limit on $B$ range between 48\% and 57\% for the same mass range. Although indirect bounds as those
from the measured rate of $b \rightarrow s \gamma$~\cite{bsgamma}
appear stronger than the results from the direct search presented
here, they can be invalidated by the presence of new physics contributions.

The interpretation of the direct measurement of the top quark mass~\cite{Amsler:2008zzb}, has become
a subject of intense discussion  in terms of its
renormalization scheme~\cite{hoang}.
The extraction of this parameter from the 
measured cross section provides complementary
information, with different sensitivity to theoretical and
experimental uncertainties, relative to 
direct methods that rely on kinematic details of the top quark reconstruction.
Simulated samples of \ttbar\ events generated at different values of the top quark mass are used to estimate the signal acceptance. The resulting measurements of  \sigmatt\  are fitted as a function of
$m_{t}$~\cite{cacciari}:
\begin{equation} \label{eq_massfit}
\sigma_{t\bar{t}}(m_t) = \frac{1}{m^4_{t}} [ a + b (m_{t} -m_0) + c
(m_{t} -m_0)^2 + d  (m_{t} -m_0)^3]
\end{equation} 
where $\sigma_{t\bar{t}}$ and $m_{t}$ are in~pb and~GeV/$c^2$, respectively, and $m_0 = 170$~GeV/$c^2$~\cite{massfit}.
\begin{figure}[t]
\centering
\includegraphics[width=0.45\textwidth,clip=]{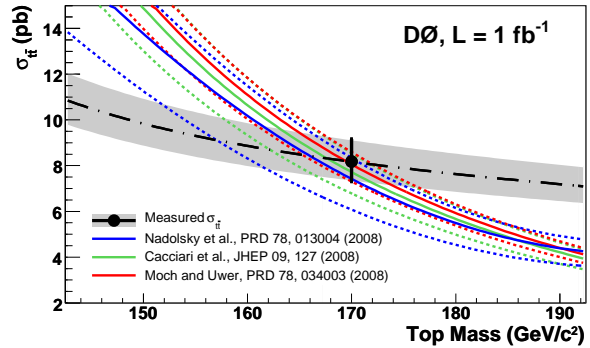}
\caption{Experimental and theoretical~\cite{nadolsky,cacciari,moch} \sigmatt\ as function of $m_{t}$. The colored dashed lines 
represent the theoretical uncertainties due to the choice of the PDF
and the renormalization and factorization scales.
The point shows the measured combined  \sigmatt, the black dashed line the fit with Eq.~\ref{eq_massfit} and the gray 
band the corresponding total experimental uncertainty.}
\label{fig:mass}
\end{figure}
The dependence on the top mass is due to the
mass dependence of the selection efficiencies. 

We compare this parameterization to a prediction in pure next-to-leading-order (NLO) QCD~\cite{nadolsky},
to a calculation including NLO QCD and all higher-order soft-gluon resummations in next-to-leading
logarithms (NLL)~\cite{cacciari},
to an approximation to the next-to-next-to-leading-order (NNLO) QCD cross section that includes all
next-to-next-to-leading logarithms (NNLL) relevant in NNLO QCD~\cite{moch},
and to a calculation that employs full kinematics in the double differential
cross section beyond NLL
using the soft anomalous dimension matrix to calculate the soft-gluon
contributions
at NNLO~\cite{kidonakis}.
Figure \ref{fig:mass} shows the experimental and the theoretical~\cite{nadolsky,cacciari,moch}
\ttbar\ cross sections as a function of the top quark mass.

Following the method of Refs.~\cite{ljets,dilepton}, we extract the most probable top quark
mass values and the  68\% CL band. Since the theoretical predictions are performed in the pole mass scheme, this defines the extracted
parameter here. The results are given in Table~\ref{tab:mass}.  
All values are in good 
agreement with the current world average 
of $171.2 \pm 2.1$~GeV/$c^2$~\cite{Amsler:2008zzb}.
\begin{table}[ht]
\begin{center}
\caption{Top quark mass with 68\% CL region for different
theoretical predictions of \sigmatt. Combined experimental
and theoretical uncertainties are shown.}
\vspace{0.2cm}
\begin{tabular}{c|c} 
\hline \hline
Theoretical prediction & $m_{t}$ (GeV/$c^2$) \\[2pt] \hline \\[-8pt]
NLO \cite{nadolsky} & $\mnlo \ermnlo$ \\[2pt]
NLO+NLL \cite{cacciari} & $\mtc \ermtc$ \\[2pt]
approximate NNLO \cite{moch} & $\mtm \ermtm$ \\ [2pt]
approximate NNLO \cite{kidonakis} & $\mtk \ermtk$ \\ [1pt] \hline \hline
\end{tabular}
\label{tab:mass}
\end{center}
\end{table}

In summary, we have combined the \ttbar ~cross section measurements  
in \ljets, $\ell \ell$ and \ltau\ channels to measure 
$\sigma_{t\bar{t}}=8.18^{+0.98}_{-0.87}$~pb 
for a top quark mass of 170~GeV/$c^2$. For the first time, we have also calculated ratios of cross sections and 
interpreted them in terms of limits on non-standard  model top quark 
decays into a charged Higgs boson. All results are in good agreement
with the SM expectations. Finally, using different theoretical predictions given in the  pole mass scheme, we have extracted the top quark mass from the
combined \sigmatt\ and have found the result to be consistent with the world
average top quark mass~\cite{Amsler:2008zzb} from direct measurements.

%
We thank the staffs at Fermilab and collaborating institutions, 
and acknowledge support from the 
DOE and NSF (USA);
CEA and CNRS/IN2P3 (France);
FASI, Rosatom and RFBR (Russia);
CNPq, FAPERJ, FAPESP and FUNDUNESP (Brazil);
DAE and DST (India);
Colciencias (Colombia);
CONACyT (Mexico);
KRF and KOSEF (Korea);
CONICET and UBACyT (Argentina);
FOM (The Netherlands);
STFC and the Royal Society (United Kingdom);
MSMT and GACR (Czech Republic);
CRC Program, CFI, NSERC and WestGrid Project (Canada);
BMBF and DFG (Germany);
SFI (Ireland);
The Swedish Research Council (Sweden);
CAS and CNSF (China);
and the
Alexander von Humboldt Foundation (Germany).


\begin{thebibliography}{99}
\bibitem{massfit} We obtain $a = 6.82350 \times 10^{9}, b = 1.10480 \times 10^{8}, c = 8.80552 \times 10^{5}$ and $d = - 1.767 \times 10^{3}$ for Eq.~\ref{eq_massfit}.
\end{thebibliography}
\end{document}